\documentclass{article}

\begin{document}

\title{Handling Pandemic-Scale Cyber Threats: Lessons from COVID-19}

\author{Adam Shostack$^1$ \and Josiah Dykstra$^2$}
\date{%
    $^1$Shostack + Associates \\%
    $^2$Trail of Bits \\[2ex]%
    \today
}

\maketitle

\begin{abstract}
The devastating health, societal, and economic impacts of the COVID-19 pandemic illuminate potential dangers of unpreparedness for catastrophic pandemic-scale cyber events. While the nature of these threats differs, the responses to COVID-19 illustrate valuable lessons that can guide preparation and response to cyber events. Drawing on the critical role of collaboration and pre-defined roles in pandemic response, we emphasize the need for developing similar doctrine and skill sets for cyber threats. We provide a framework for action by presenting the characteristics of a pandemic-scale cyber event and differentiating it from smaller-scale incidents the world has previously experienced. The framework is focused on the United States. We analyze six critical lessons from COVID-19, outlining key considerations for successful preparedness, acknowledging the limitations of the pandemic metaphor, and offering actionable steps for developing a robust cyber defense playbook. By learning from COVID-19, government agencies, private sector, cybersecurity professionals, academic researchers, and policy makers can build proactive strategies that safeguard critical infrastructure, minimize economic damage, and ensure societal resilience in the face of future cyber events.
\end{abstract}

\section{Introduction}

The COVID-19 pandemic forced humanity to confront a widespread, deadly, and rapidly spreading threat. It produced horrific impacts to human life and the global economy. While there are notable differences from the biological world, COVID-19 offers an opportunity to consider how we should think about preparation and response to catastrophic digital events. Human, process, and technological systems in 2024 are unprepared for pandemic-scale digital threats but timely and effective responses are more likely if we incorporate lessons from COVID-19.

COVID-19 disrupted life and societies in ways we all hope are once in a lifetime experiences. People had to re-tool their daily routines and businesses had to pivot. Having kids at home, working remotely, and being unable to visit with vulnerable friends and family impacted us all in significant ways. Many people in essential jobs or service industries had to continue working in-person through rapidly changing rules, and they were often the public face of evolving guidance to a confused, scared, and angry public.

Numerous reports have documented how COVID-19 produced second-order effects in the form of new cyber threats and also impacted the practices of cyber defense. These included: stress to migrate to remote-mostly work, the stress of maintaining cybersecurity in resource-constrained organizations, cyber threats to healthcare research and institutions~\cite{he2021health}, and phishing and misinformation fueled by COVID-related themes~\cite{lallie2021cyber}. Though not the focus of our work, these second-order effects are significant and  demand further study and mitigations.

We take a complementary view, exploring how the response to a disruptive global cyber threat could be informed by responses to the biological COVID-19. Responding to a pandemic-scale cyber event would require international attention and cooperation, and would benefit from effective preparation. However, for the sake of scope and as a starting place, we intentionally limit our focus specifically on considerations at the national level in the United States. We acknowledge that certain types of cyber events have geographic components, that closing a border is more complicated in cyberspace than in the physical world, and that responses to both pandemics and cyber events tend to be organized by governments where international coordination is important. Future work must consider these issues.

Pandemic-scale events (PSE), defined in Section~\ref{section:pse}, may seem too big, too rare, or too complex for which to have a tangible, useful plan and playbook. Nevertheless, this is not a ``what if?'' proposal but one that asks: ``what when?'' ~\cite{herley2015if}. There is, in reality, more variability to potential PSEs than to the similarities that bind biological pandemics. The field of threat modeling offers helpful insights for thinking about PSE. By proactively outlining potential scenarios, decision points, and mitigation strategies, even a high-level playbook can significantly reduce the chaos and improve response times during a large-scale cyber event. This preparedness fosters a more measured and coordinated defense, minimizing potential disruption and damage. Threat modeling, with its focus on understanding systems and what can go wrong with them, provides a structured approach to building this playbook. By applying these activities and perspectives to a PSE, we can anticipate groups of problems, predict potential consequences, and develop response plans that effectively address the unique challenges of a cyber pandemic.

If done correctly, preparing the United States and other nations for a pandemic-scale cyber event would be powerfully beneficial. Preparation and practice are key to avoiding costly mistakes---economic and otherwise---from action bias and knee-jerk decisions during a crisis~\cite{dykstra2022action}. Preventative response strategies also support the defense and resilience of national critical functions during a pandemic-scale cyber event. Collaboration will be key, just as it is key to life and safety in critical functions from transportation to food supply. Modern cybersecurity is more than technology alone: it is also about developing doctrine, skills, and role assignments. Nowhere is this more clear than in pandemic-scale cyber events.

We present the case for immediate and comprehensive preparations for a pandemic-scale cyber threat using extensive lessons learned from COVID-19. In Section~\ref{section:pse} we define and characterize a pandemic-scale event. In Section~\ref{section:covid19}, we draw six critical lessons from COVID-19 that are instructive for cybersecurity. Section~\ref{section:cybersecurityIssues} presents critical considerations. We offer ideas to aid an actionable playbook in Section~\ref{section:playbook} and conclude in Section~\ref{section:conclusion}.

\section{Pandemic-Scale Cyber Events}
\label{section:pse}

In this paper, we consider events that require every member of a large group to change their behavior, resulting in different ways of living or doing business for a month or more. Historically, such events are infrequent. Therefore, we focus our attention on `every few decades' events and lay out what those might be. In the language of public health, the defining distinctions between epidemics and pandemics are the speed and scale of the spread of an infectious disease. Digital threats can be both communicable and noncommunicable, yet both can affect large groups and require behavior change. Such cyber events could be vulnerabilities in hardware and software, but they could also result from and involve multi-domain and multi-discipline issues such as large-scale physical world phenomena in the form of a Carrington Event~\cite{tsurutani2003extreme} or a large electromagnetic pulse. Despite the low frequency of occurrence of severe geomagnetic storms is ``the potential for permanent damage that could lead to extraordinarily long restoration times''~\cite{NAP12507}.

Pandemic-scale events are more than extraordinary vulnerabilities or weaknesses. Vulnerabilities such as Heartbleed or the collection of Log4Shell issues are indeed ``all hands on deck'' events that stress a segment of the digital system, including business owners and cybersecurity professionals. However, those vulnerabilities are regular, going back to the so-called ``summer of worms'' or the ``Sendmail bug of the month club.'' Slammer and Code Red, while impactful, did not require sustained behavioral change for a broad group of users (though one could argue that staff at Microsoft, whose development processes changed, meet these criteria). These vulnerabilities shared characteristics of being in internet-facing software, often in pre-authorization code, and being easily exploitable. In this sense, they are comparable to forest fires or floods. Emergency response professionals have plenty of experience and deterministic processes for dealing with them. Such events may dramatically upend or even end lives, but they do so in constrained ways. We can also say that events such as the TJX hack of credit cards, the Equifax leak of financial information, the Mirai botnet DDoS of Dyn DNS, or the OPM hack did not require responses of pandemic scale or duration.

One can imagine widespread cyber threats that do require such a large-scale response, either behavioral change or manufacturing new equipment. For example, a more easily exploited version of Spectre or Meltdown forcing customers from the cloud en masse and for an extended time might require designing, manufacturing, delivering, and deploying new hardware at cloud providers. The Gigabyte or Barracuda incidents of 2023 may have been such incidents. Gigabyte had a motherboard design flaw in which motherboards were insecurely downloading and executing content. This was discovered in 2023, and Gigabyte was able to issue a patch. Barracuda, whose flaw was also discovered in 2023, actually recommended replacing their email security hardware appliances. What if the Gigabyte flaw was not as easily patched? What if the Barracuda issue impacted, say, Cisco or Juniper, or other vendors of core internet routers? Those routers are exposed to the internet, and replacing thousands of them could exhaust inventory and overwhelm lean manufacturing that's intended to deliver a handful of large routers daily. It would likely drain the supply chain of various components, limiting production and impacting the production of other goods. (Availability of chips reduced car-making capacity by an estimated 9.5 million vehicles during the pandemic, while floods in Thailand in 2011 threatened to slow production of PCs globally~\cite{brinley2024the,arthur2011thailand}). Although some parts of the internet are resilient, others are fragile. The power grid is vulnerable, and an outage would almost certainly be catastrophic. Likewise, a long-term DNS outage would hamper recovery (along with other activity). The Crowdstrike outage of July 2024 has been described as impacting 674,000 organizations~\cite{interos}.\footnote{The cited report is the largest claim of impacted systems we could find. We do not endorse their analysis but find it useful as a reminder of the apparent scale of the incident.} Despite the scale of the problem, we do not believe it represents what we are calling pandemic-scale because, for most people, it was `follow these steps, once.' Similarly, other very large incidents such as Solarwinds and OPM did not involve behavior changes from a large group.

Others are also worried, including the financial sector. The 2008 financial crisis was one example of systemic risk producing severe outcomes and led to dealing with institutions that were ``too big to fail.'' In 2021, Federal Reserve Chairman Jerome Powell said that ``the risk that we keep our eyes on the most now is cyber risk''~\cite{fanti2022missing}. A cyber attack against a single financial organization could spread and magnify into a pandemic-scale event.

Thankfully, we assert that there has not yet been a cyber threat comparable to COVID-19 in its scale and impact. While such cyber impacts are conceivable, we do not judge their likelihood. If one takes place, then we will wish to be prepared, and here we can learn lessons from Covid preparation and response. 

We are not the first to explore catastrophic cyber incidents. One 2007 study urged the cybersecurity community to prepare for the ``propagation of widespread failure or malfunctioning in critical infrastructure systems with an associated large quantum of harm to society'' ~\cite{michael2009cyberpandemics}. They presented characteristics, signs, and symptoms without concrete proposals for how to prevent, prepare for, and respond to such events. The World Economic Forum has fostered conversations on the idea of a ``cyber pandemic'' ~\cite{Kaye_2021}. Eita and Gazit~\cite{eitan2023no} examined a six-hour Facebook outage that occurred in 2021, offering a microcosm of the potential effects of a wide-scale outage. Survey respondents reported a complex mixture of responses, including even joy of being disconnected for a short time. In a pre-Covid scenario-based study of 754 U.S. students, participants predicted the impacts of an extended and widespread internet outage and focused more on personal impacts than societal impacts~\cite{grandhi2020internet}. We extend and expand on these works.

\subsection{Characteristics of a pandemic-scale event}
The classical definition of a pandemic is ``an epidemic occurring worldwide, or over a very wide area, crossing international boundaries and usually affecting a large number of people''~\cite{porta2014dictionary}. This definition is insufficient when applied to cyber events because scale is not the only meaningful and distinguishing characteristic. There might be an urge to characterize a PSE by an order of magnitude scale, such as 10x or 100x more victims, more days of outage, or more financial loss than some reference event. History shows that larger and larger data breaches have not induced changes in behavior, and it is unclear if there is a scale at which they would. Most customers of the victims of data breach make no change to their behavior~\cite{janakiraman2018effect}.

We characterize pandemic-scale events as those that require every member of a large group to change their behavior, resulting in different ways of living or doing business for a month or more, or events that violate long-standing assumptions. This section increases the specificity of these criteria. We refer to these variously as pandemic-scale events (PSE) and mean the same thing by the slight variants that are demanded by clear writing. 

\textbf{``Every member of a large group.''} If we treat this as ``the security team cannot clean this up on its own'' then this criterion is the most commonly met, as many issues require the involvement of technology operations. Replacing a motherboard might be complex for every employee; moving from the cloud might entail a crash cross-company effort.

\textbf{``Behavior change.''} The change is bigger than ``follow these steps once.'' For example, a flaw in Bluetooth might entail turning off all Bluetooth devices, including on Apple devices, AirDrop, AirPlay, and ``find my'' including AirTags, and requires either new ways of achieving those goals, or accepting that they are unachievable for some time.

\textbf{``New Manufacturing.''} An issue that permanently damaged hardware could take months or years to recover from. We have seen harbingers of such issues from the Pentium 4 math flaw to Spectre to BMC/IPMI flaws.

\textbf{``Assumption-violating.''} Pandemic planners assumed that transmissibility would be accompanied by symptoms, such as fever or cough. Those would enable rapid screening at airports and elsewhere to slow initial spread. Covid violated that assumption of ``no symptom-free transmission.'' Cybersecurity may have equivalents. The most obvious would be a failure of a major cryptographic algorithm, such as SHA-2-256 or mode of operations, such as AES-GCM or AES-CCM. New vulnerability families with widespread instances that are easily found (such as the organized recognition of format string issues) might fit, but we note both cryptographic failures and new families of vulnerabilities happen ``every few years.''

Other researchers have compared the attributes of cyber and biological threats; instead, our aim is to improve preparations for future cyber incidents using observations from the global response to COVID-19. The common attributes we consider definitive are: threats that are novel, highly transmissible, pose serious complications, and have rapid global reach. Even ransomware and social engineering, while significant, do not reach this threshold. However, it is possible to imagine malware that meets these four criteria.

As an aside, activities of daily living (ADL) is a term that has been used in healthcare since 1950 to refer to the basic tasks of everyday life that individuals do independently and which contribute to safety and quality of life~\cite{edemekong2019activities}. ADL includes eating, dressing, and mobility. Researchers have explored the impact of pandemics on ADL by studying changes in individuals' ability to perform essential tasks due to various factors related to pandemics and other causes. It would be interesting to explore the items that constitute \textit{digital} activities of daily living, especially if a scale or index could quantify the impact of a PSE.

\subsection{Limits of the pandemic metaphor}
\label{section:limits}
There are places where the digital world is meaningfully different---even better prepared---than biology. Vulnerabilities are regularly patched. The largest software makers already have well-oiled patch distribution mechanisms in place, which may or may not function in a PSE. Harm can also be constrained in some cases. Attacks that leverage trust infrastructure may be limited, even to a single victim. In cases such as the North Korean attack on Sony or the Iranian attack on Saudi Aramco were highly destructive, but the destruction was constrained. In the Aramco case, 30,000 PCs were wiped and needed to be re-installed, but they were re-installable. If their UEFI had been overwritten, Aramco could have needed to replace 30,000 machines.

Biological diseases often sicken or kill their hosts and are further specifically dangerous to caregivers who breathe the same air, are exposed to other effluvium, may be exposed to blood, and so forth. People who recover from a disease may have immunity to further infection, although pathogens do mutate. Biological recovery may also be more of a spectrum than the effectively binary ``infected or not'' digital world, such as Long COVID or post-polio syndrome.

There are places where the digital world is more vulnerable than the biological world. Since many devices are networked by design, for instance, the speed and reachability for threats to propagate are tremendous. Even twenty years ago, Robert Morris could say, ``To a first approximation, every computer is directly connected to every other computer in the world.'' The three months of warning we had as Covid emerged may not be available to us during a cyber event.

Finally, the speed and scale of a health or societal issue can be categorized as a pandemic, epidemic, or even a ``slow epidemic.''
Pandemics are characterized based on the rate of spread and geographic scope~\cite{columbiasph}. There are also epidemics, and a concept of `slow epidemics' which emerge over many years. 
In 1997, for example, the World Health Organization declared obesity a global epidemic given its rapid and widespread increase in the preceding decades~\cite{world2000obesity}.
Similarly, in response to the rise of phishing, we have slowly accepted the inconvenience of multi-factor authentication, the false-positives of spam filters, and other changes to all our behavior. Had the onset of phishing been sudden, perhaps it would be a pandemic-scale event.

\section{Lessons for Cyber from the COVID-19 Pandemic}
\label{section:covid19}

In their 2023 book, \textit{Lessons From the Covid War}, the Covid Crisis Group documented first-hand experiences, numerous interviews with government and industry stakeholders, and expert analysis of key themes and challenges by looking back on the past three years~\cite{covid2023lessons}. The Group included distinguished professors, authors, historians, CEOs, and former government executives. They self-assembled to lay the foundation for a National Covid Commission patterned after the 9/11 Commission, but the U.S. has never assembled such a Covid Commission. Because these experts identified core lessons learned from a healthcare and public health perspective, we use their work as an informative guide for thinking about pandemic-scale cyber threats. These are not the only relevant lessons nor considerations for cybersecurity, but they illustrate fundamental topics for understanding and action today.

\subsection{Lesson 1: Trustworthy statistics are a foundational requirement} 
The detection, tracking of spread, containment, and analysis of the impact of pandemic-scale events are made possible by data reporting and institutions that gather, analyze, and report on it. That is, statistical infrastructure. This infrastructure was essential during COVID-19. Biomedical surveillance relies on infrastructure and participation that cannot be quickly established after a threat emerges. The Covid Crisis Group writes that ``there is an emerging consensus in the healthcare industry that necessary data sharing in the United States can no longer be strictly voluntary, hit or miss''~\cite{covid2023lessons}.

However, data is too often seen as ``boring'' and chronically underfunded. Data systems were creaking along when Covid arrived; some literally using faxed reports~\cite{Tahir_2020}. According to the Congressional Research Service, ``CDC has been working to transition public data surveillance to more robust integrated electronic systems \textit{for decades}; this process was incomplete when the pandemic began'' (emphasis added)~\cite{Sekar_Napili_2020}.

Official data collection analyzed exclusively by authorized professionals limits the speed and viewpoints of knowledge produced from the data. Volunteer data scientists were motivated to help spot new trends and suggest new mitigations. These lay analysts almost certainly had more time available to dive deeply and explore broadly. Crowdsourcing must be done ethically and responsibly. The legal protections afforded to health data and implications for cybersecurity have been carefully reviewed by Sedenberg ~\cite{sedenberg2015public}.

Today, the collection and sharing of cybersecurity data is highly fragmented and often compartmentalized. Technology companies have tremendous volumes of data about both their own software and infrastructure but also observations about traffic they can see on the internet. In a pandemic-scale cyber event, some consolidation and authoritative sources would be required. As the CDC has done with health data, a government entity such as CISA could be a logical hub to collect and share cybersecurity statistics. Others have proposed a focused Bureau of Cyber Statistics~\cite{king2020cyberspace}.

\subsection{Lesson 2: Informal networks are incredibly powerful}
In addition to formal relationships forged through regulation, voluntary organizations, and contractual obligations, established informal connections and new human networks emerged during COVID that produced unexpected value. The Covid Crisis Group wrote that ``Much more valuable in the crisis was simply the large informal network of scientists and doctors working in hospitals and labs around the world, including networks connected through more influential nonprofit foundations''~\cite{covid2023lessons}.

Take one specific example. ``During Operation Warp Speed\ldots Patel---a pharmacist---was a key member of the team that worked out the blueprints for how hundreds of millions of Covid vaccines could be distributed in an unprecedented partnership with private pharmacy chains like CVS and Walgreens''~\cite{covid2023lessons}. In a pandemic-scale cyber event, we should assume that some brilliant, unexpected individual contributors may rise up to be key players. We should expect unconventional solutions, perhaps from local businesses or experts, to help people deal with the cyber threat. We may also need to deal with people whose unusual and strongly held ideas make them seem like cranks. Who knows, they might be cranks or brilliant loners.

Cybersecurity professionals augment formal response activity with informal networks, from personal connections to private chat channels. Establishing trust develops over time, but in crisis, these relationships can overcome slower, formal sharing. Preparations for a cyber PSE should acknowledge that these informal networks will play a part in the response, provide data and tools to empower them and find ways to maximize their benefits. Informal networks have been acknowledged as a lesson learned as far back as the cyber defenders who responded to the Morris worm~\cite{rochlis1989microscope}. Scenario-driven tests should better consider the role of informal relationships.

\subsection{Lesson 3: Unclear roles and responsibilities inhibit progress}
\label{lesson:unclear_roles}
Healthcare delivery in the modern world is complex, involving webs of physicians in private practice, some who work for hospitals, and others who work for both. There are local and chain pharmacies, testing labs, drug researchers and producers; the list goes on and on. The interdependencies run throughout the ecosystem.

Healthcare governance in the United States is also complicated, with many agencies and no central lead. There are institutions for research (e.g., National Institutes of Health), drug safety (e.g., Food and Drug Administration), public health (e.g., Centers for Disease Control and Prevention), Centers for Medicare and Medicaid Services, healthcare (private and public), and biopharma (public and private). While federal departments and agencies can provide leadership and guidance, state and local governments ultimately play a unique role in the actual delivery of healthcare, not to mention governance. Getting vaccines into the arms of people happens at the community level.

Given the current decentralized nature of cybersecurity roles and responsibilities in the United States, there is little reason to expect that the response to a cyber PSE would be spontaneously well-coordinated. Rigidly selecting leadership is not as important as a shared understanding of who has the authority and capacity to achieve specific outcomes. As with local healthcare delivery, some pandemic-scale cyber events would require local responses, like those needing hardware replacement. For many people, even the installation or reconfiguration of software would require one-on-one assistance.

\subsection{Lesson 4: Communication must be clear and effective}
The timely creation, distribution, delivery, and consumption of useful information is essential in a crisis. Two aspects worth highlighting are the vulnerability of communication channels and the recipient's trust and confidence in the message.

\textbf{Lesson 4A: Our digital communications are vulnerable}. An enormous portion of human communication, information distribution, and entertainment is enabled by the internet today. If a pandemic-scale cyber incident disrupted internet communications, what/who would take its place? Both broadcast and personal communications are now fundamentally digital. TV, radio, and newspapers all depend on both the internet and local computerized systems for production and transmission. A PSE that renders computers or networks disabled could inhibit the widespread dissemination of information about fixes and recovery.

\textbf{Lesson 4B: Poor communication erodes trust and confidence}. Humans crave social connection and communication, and never more than during the uncertainty of a crisis. During COVID-19, people adapted to Zoom calls and outdoor gatherings to overcome isolation. When the sick and dying were unable to be with their loved ones, healthcare workers improvised and enabled communications between them. 

Effective public communication was inhibited by scientific disputes, such as droplet versus airborne transmission. Those communication difficulties were magnified by political interference and non-compliance at all levels of many governments. For example, President Trump announced that he would not wear a mask and Prime Minister Boris Johnson held parties at 10 Downing Street. The result was confusion and created a vacuum for misinformation. Politicization should be expected in the near term. 
Moreover, providing medical misinformation during Covid proved profitable and provided a path to prominence. For example, the Washington Post has documented how ``four major nonprofits that rose to prominence during the coronavirus pandemic by capitalizing on the spread of medical misinformation collectively gained more than \$118 million between 2020 and 2022''~\cite{weber2024tax}. It seems reasonable to consider the possibility of misinformation in the wake of a PSE. Cyber information sharing groups that arose during the pandemic were attacked for ``censorship''~\cite{rogers2024statement}.
Establishing facts---and getting people to understand and accept them---is a problem beyond pandemics alone; however, preparation and testing of effective communication must be developed and tested now.

Communication plays a pivotal role in building human resilience by facilitating understanding, fostering community, and promoting effective responses to challenges. 
One analysis of lessons from the pandemic emphasized the theme of resilience. The authors highlight ``clear and accessible communication'' as essential to crisis management, no matter the threat~\cite{mott2023preparing}. Additionally, they point out that it was helpful that communication and data analysis weren't done solely by governments; the public, academics, and industry were encouraged to use available data to aid response efforts.

\subsection{Lesson 5: Existing plans may be insufficient to guide action}
\label{lesson5}
Even before COVID-19, pandemics were a serious enough threat that playbooks existed for them. Unfortunately, as the Covid Crisis Group discusses, ``\ldots the playbook did not actually diagram any plays. There was no `how.' It did not explain what to do''~\cite{covid2023lessons}. The plans were insufficient to guide action, and this was magnified by the non-symptomatic manifestation of Covid in many victims.

Existing plans can be ineffective if underlying assumptions are violated or facts change. For example, a pandemic playbook may assume that citizens will largely trust and follow medical science. The unexpected scale of misinformation and disinformation sharply reduced the uptake of common disease precautions, including masking, distancing, and vaccinations. Next time may be worse. Disinformation tools are growing more powerful. There's a possibility that a PSE will be related to a cyber-attack executed with ``combined arms'' of a technical attack and an associated disinformation campaign.

The unfolding of a pandemic-scale cyber event will require diagnosis but also what to actually do. In 2021, Executive Order 14028 on Improving the Nation's Cybersecurity included a section ordering ``Standardizing the Federal Government's Playbook for Responding to Cybersecurity Vulnerabilities and Incidents.'' Implementation of this goal is fragmented. The Cybersecurity and Infrastructure Security Agency (CISA), for example, has the lead for Federal Civilian Executive Branch (FCEB) information systems. It is unclear whether or not CISA would have the lead for a response to a vulnerability in every smartphone or router.

Finally, plans and preparation must be combined with the ability to be agile. President Eisenhower famously advocated that the practice of developing plans and exploring all the options was of greatest value. In the early days of COVID-19, it was unclear how the virus was spread. As a result, early mitigations such as cloth masks were later found to be ineffective. As Mike Tyson once said, ``Everyone has a plan until they get punched in the mouth.''

\subsection{Lesson 6: Syndemic issues will likely inhibit response}
Syndemics (concurrent or sequential epidemics) can occur when social, political, or economic factors magnify impact or inhibit responses~\cite{singer1992generations}. Covid and other viruses are exacerbated by pre-existing societal issues and health disparity with a disproportionate impact on vulnerable populations. In the digital realm, there will also be greater disadvantages for groups and organizations disadvantaged by a lack of resources, awareness, and knowledge. The digital divide mirrors inequities in the physical world, and marginalized groups may be especially affected by a PSE if they are not considered explicitly.

The concept of syndemics can be applied to the realm of cybersecurity. The digital world is more and more interdependent every day. This complexity makes cascading failures increasingly likely and also difficult to map. Vulnerabilities within a system can cluster, amplifying the impact of individual exploits. For instance, unpatched software with multiple vulnerabilities might create a domino effect if one flaw is breached. Even redundancy, a typically desirable approach to resiliency, could inadvertently lead to cascading failures.

Social factors can also play a syndemic role. A phishing attack targeting a population already stressed by an external crisis (economic downturn, natural disaster) may have a higher success rate due to heightened anxieties. Furthermore, a lack of cybersecurity awareness within a user base can hinder effective response to an attack. Understanding these syndemic dynamics in cybersecurity is crucial for developing robust defense strategies. By acknowledging the interplay between technical vulnerabilities and social-behavioral factors, organizations can improve their preparedness and response capabilities.

Note that syndemics can also arise from the cumulative effects arising from a sequential epidemic and lead to other widespread changes. In Section~\ref{section:pse}, we noted that the internet worms of the early 2000s---Code Red (2001), Nimda (2001), Blaster (2003), Slammer (2003)---were not pandemic-scale events themselves; however, they did lead to a recognition and fundamental shift in secure software development practices and patch management.

\section{Additional Confounding Cybersecurity Issues}
\label{section:cybersecurityIssues}

The lessons from COVID-19 point a spotlight on areas of cybersecurity that contribute to vulnerability in the preparation for a PSE. In this section, we provide additional factors that make PSE readiness and response particularly challenging today.

\subsection{Vital statistics and data infrastructure are missing}
The public health system has data gathering systems, including criteria for what symptoms constitute a disease and mechanisms to routinely collect, summarize, and distribute such data. Data is gathered around a variety of contact points with the healthcare system, including hospital admissions, deaths, and ``reportable diseases,'' those of great concern to the public health system. The responsibility for reporting is imposed on health-care providers, usually as an ethical requirement and a requirement of licensing.

These became controversial in COVID-19 when disputes arose over whether or not people die ``of Covid'' or ``with Covid.'' There is ongoing work to measure ``excess deaths'' relative to pre-Covid levels. 

Thus far, there is no institution gathering and publishing public vital statistics for cybersecurity that would give insights about PSE vulnerability and resilience. It is challenging (even for Microsoft) to know precisely the number of Windows computers on the planet, nor what percentage of them are patched. Vendors, such as Cisco, are also unlikely to reveal the number of high-end routers they could produce in a month.
We even lack shared terminology, such as an equivalent of deaths, except for the rough approximation of ``bricking'' devices or other denial of service attacks~\cite{Shostak_2022}. The United States lacks mandatory reporting outside of regulated industries, and we lack statistical bodies to bring it all together. The limited data collected privately, such as Verizon's annual Data Breach Investigations Report, can be a useful snapshot but is narrowly focused and not helpful in real-time health surveillance.

\subsection{Roles and responsibilities are overlapping and complex}

In Section~\ref{lesson:unclear_roles}, the Covid Crisis Group highlighted consequences to Covid response from unclear governance. The world of cybersecurity is no less complex than health. Authorities and regulations distinguish roles for policy and strategy, not to mention commercial hardware and software vendors, cybersecurity companies, etc. The front lines are dominated by industry, with few state and local resources for cyber threats. 

American responses to many areas of cybersecurity already exhibit organizational complexities similar to Covid. The United States government has the Office of the National Cyber Director, responsible for policy and strategy, which is distinct from law enforcement (e.g., FBI), foreign intelligence (e.g., NSA), domestic guidance (e.g., CISA), standards (e.g., NIST), and defense (DoD). The Federal government regulates and responds to cybersecurity 
sectorially, for example:
\begin{itemize}
    \item Food and Drug Administration (FDA) issues guidance for device makers on safety and effectiveness of medical devices.
    \item Securities and Exchange Commission (SEC) issues guidance for publicly traded companies.
    \item Transportation Security Administration (TSA) issuing mandatory rules for pipelines and more.
    \item Department of Defense (DoD) issues mandatory rules for the defense industrial base.
    \item Banks are regulated by the Federal Financial Institutions Examination Council (FFIEC).
\end{itemize}

Cybersecurity is also regulated ``horizontally'' across departments and agencies at the Federal level, including:
\begin{itemize}
    \item The Federal Trade Commission (FTC) penalizes companies for ``unfair'' or ``deceptive'' practice.
    \item Under Executive Order 14028, companies selling to the U.S. Government must attest to security practices in accordance with the NIST Secure Software Development Framework (SSDF).
    \item CISA has issued guidance encouraging companies to ship products that are secure by default and secure by design.
\end{itemize}

State regulations include a mix of both sectoral and general regulations, and many of the privacy regulations target any company doing business in their state. For example:
\begin{itemize}
    \item California Privacy Protection Act and Virginia Act for Protection of Personal data (and roughly a dozen laws which follow the Virginia model, lacking a ``right of private action'' in the California law).
    \item New York's Department of Financial Services regulates banks licensed in New York, meaning all the significant ones.
    \item Washington State's ``My Health, My Data'' recognizes that a great deal of data can be used to infer information about gender and reproductive health, and regulates it accordingly.
\end{itemize}

In non-crisis times, sector-specific cyber regulation is desirable because it tailors communications, defenses, and plans to those most relevant to a subgroup. This segmentation, unfortunately, becomes a liability to broad-scale coordination unless barriers can be quickly lowered.

\subsection{There are no plans for a pandemic-scale cyber event}
In Section~\ref{lesson5}, a lesson from Covid was that existing plans were insufficient. To the best of our knowledge, there is \textit{no} U.S. government or industry plan for an event of the scale we describe as ``pandemic-scale.'' Dan Geer has defined security as the absence of unmitigable surprise~\cite{geer2022extracting}. The events we describe here are more difficult to mitigate and thus security requires not winging a response.

In 2018, researchers presented a postmortem of a hypothetical North American blackout from a cyber attack in 2038~\cite{anantharaman2018going}. This paper addressed the technical vulnerabilities and shortcomings but was silent about human, societal, and governmental aspects of response and responsibilities to the scenario. Expanded scenarios like this could be instructive for planning responses to pandemic-scale events.

One construct that exists today is trust communities, including computer emergency response teams (CERTs) and computer incident response teams (CIRTs). At national (US-CERT), sector, and organizational levels, these entities are designed to coordinate defenses and respond to cyber incidents. The CERT Coordination Center (CERT/CC) was first created to respond to the Morris worm. The relationships and partnerships that exist within and between emergency response organizations could undoubtedly play a role in responding to a PSE.

Finally, a reviewer pointed out that ``There's engineering, and there's politics.'' Some work, like strengthening trust networks and encouraging people to develop contact methods that would survive Slack or Discord being offline, is engineering~\cite{leveson2011engineering}. Some of the systemic work may be political.

\subsection{Example: election security as illustrative of problems from a PSE}

To help show the complexity of pandemic-scale cyber events, we offer election security as a real-world non-PSE microcosm of the issues facing the United States.

In the U.S. Federal system, elections are run by the states. As a result, the availability and quality of cybersecurity training, resources, and capabilities vary greatly. For example, in Missouri, there are approximately 80 local election authorities, and their emails are handled by a variety of .com, .org and even a few .mo.gov email domains. If Missouri chooses not to prioritize cybersecurity, or a problem arises despite their best efforts, that is their right and responsibility. So, for example, Missouri must defend its elections against threats from Russia cyber attacks, and the need for election offices to operate securely has been the subject of a great deal of disinformation. There are certainly some resources available from the Secretary of State, from CISA, and from a variety of non-profit organizations, but the resources and potential incident response are decentralized, involving organizations with many different authorities, reporting structures, mandates, and budgets.

Is this ``a change that would require every member of a large group to change their behavior for a month or more?'' Perhaps not. It might be on the borderline for those who live in vote-by-mail states and less for those who live in locales with in-person voting. However, the challenges we face in handling election security are illustrative of the challenges we would face in a PSE.

Cyber attacks against election infrastructure, and the very real disinformation campaigns waged against elections in general, can inform what might happen in a PSE.

\section{Towards a Playbook for Pandemic-Scale Cyber Events}
\label{section:playbook}
We urgently need a PSE playbook. A first national PSE playbook could be developed by a small team with a few months of work. The United States does a great deal of disaster readiness planning in other domains and can build on that experience. The development, rehearsal, and implementation of playbooks are intended to speed the response time and effectiveness when situations arise. Good playbooks are actionable, meaning they provide tactical actions and assignments of tasks. They should also be tested to ensure they are realistic. Playbooks are aided by data collection that enables forward warning indicators and crisp criteria for execution.

\subsection{The development of playbooks}

There is no known playbook for developing playbooks, despite their prevalence. In cybersecurity, the concept of playbooks evolved gradually as the field matured. Playbooks emerged because cyber events necessitated more structured and rapid response and the need for a methodical approach. Although various plans, playbooks, and government organizations exist today, they are neither individually nor collectively sufficient for a PSE. Importantly, a PSE is an extraordinary incident but general incident response playbooks are insufficient for the complexity and impact of a PSE.

When it comes to formalizing and standardizing the concept within the context of cybersecurity, the National Institute of Standards and Technology (NIST) has played a significant role through its guidelines and frameworks.
NIST's work has significantly influenced how organizations manage cybersecurity risks, respond to incidents, and implement security protocols. It has been instrumental in the widespread adoption and refinement of playbook concepts through its publications.
One of the foundational documents that could be considered as aligning with the concept of a cyber playbook is NIST Special Publication 800-61, ``Computer Security Incident Handling Guide''~\cite{cichonski2012computer}. First published in 2004 and revised several times since, SP 800-61 provides guidance on handling and responding to computer security incidents. 

CISA has a \textit{Federal Government Cybersecurity
Incident \& Vulnerability Response
Playbook} that incorporates several NIST references~\cite{cisa2021federal}. However, this playbook does not discuss cyber events on a pandemic scale. The focus of this playbook is on incidents affecting Federal Civilian Executive Branch (FCEB) systems. Pandemic-scale cyber events may not be limited to FCEB systems and could involve a wider range of threats and stakeholders. There is also a \textit{National Cyber Incident Response Plan} (NCIRP) specifically for ``significant cyber incidents posing risks to critical infrastructure''~\cite{dhs2023national}. It is explicitly not a tactical or operational plan.

The Federal Emergency Management Agency (FEMA), likewise, published ``Planning Considerations for Cyber Incidents: Guidance for Emergency Managers''~\cite{fema2023planning}. A PSE may qualify as an emergency under FEMA's definition, and this document does describe roles and responsibilities for emergency managers at the state and local level to aid safety and economic impacts on individual communities. This is perhaps the closest to considering the unforeseen and far-reaching consequences of PSE.

Newly established in 2023 is the Office of Pandemic Preparedness and Response Policy (OPPR) within the Executive Office of the President (EOP). It is tasked to lead preparedness for, and response to, biological threats that could lead to a pandemic in the United States. Also within EOP is the Office of the National Cyber Director (ONCD). Alone, or together with OPPR, ONCD is a natural lead for developing the national-level PSE playbook.

\subsection{Key playbook questions}
Traditional, non-PSE playbooks exist in various forms across cybersecurity today, often inside individual companies. PSE playbooks will look quite different and cannot simply be scaled-up versions of corporate playbooks. The development of PSE playbooks will require the input of multiple stakeholders and should be developed through workshops with experts from government, industry, and academia. Then, the plans must be regularly tested with exercises to ensure that they are effective and that stakeholders gain experience with their execution.

Developing PSE playbooks will require the input of many stakeholders and demand evolution as we continue to gain knowledge and experience. Rather than offering an insufficient draft here, we instead propose the following questions and considerations for the future development of a PSE playbook:

\begin{itemize}
    \item How do we empower informal networks? Is there value, for example, in subsidizing community conferences (e.g., B-Sides) or ensuring that attendee lists are available? Should there be a Good Samaritan exemption to NDAs? For example, should we shield Amazon employees from consequences for sharing information in a crisis? Shield government employees who gave out useful information?
    \item How do we enable and empower people closest to those who need help? For example, what might be the role of a local Best Buy in St. Louis in aiding local residents?
    \item What mechanisms might we want for crowdsourcing and enabling crowdsourcing with infrastructure in advance? 
    \item How do we track and manage known risks (similar to the risk of not being able to make enough PPE)?
    \item How do we track emergent risks? Is there an equivalent of 20 virus families? Geographic?
    \item How do we accelerate and optimize response if there is a problem?
    \item What's the role for CERTs/CIRTs/ISAC/ISAO and other trust communities in pandemic-scale response? How can their experience and networks be applied effectively?
    \item What early warning indicators already exist?
    \item How could we stress test or run a reasonable tabletop exercise?
    \item What privacy implications must we consider?
    \item How do we protect the ``caregivers'' and other front-line workers who need to deliver the PSE response, including keeping their devices, networks, and equipment from infection or destruction?
    \item What does recovery look like? Is there a ``return to normal'' or ``adjust to new normal'' phase?
\end{itemize}

\subsection{Playbooks as necessary but not sufficient}
Pandemic playbooks offer a warning: while they existed before Covid, in retrospect they were unfortunately not actionable enough to be relevant or useful. This error can be avoided by regularly updating and stress testing them as realistically as possible.

More is needed than a playbook document alone. To be effective, they need regular testing, evaluation, and updating as organizations and technology change. Playbooks require agreement on roles and responsibilities among the stakeholders. One research assessment of incident response playbooks found that some lacked sufficient detail for real-world use, particularly for junior staff~\cite{stevens2022ready}. Their recommendations are likely to be even more important for the urgency, impact, and complexity of a pandemic-scale event.

Between 2006-2024, DHS  conducted nine national cyber exercises, known as Cyber Storm, focused on testing national cybersecurity guidance and federal roles. Among the key findings in 2022 was that ``During the exercise, national plans and policies as written and discussed had limited impact or influence on
private sector response''~\cite{cisa2022cyber}. Overall, the lessons learned supported our assertion that there is no PSE playbook and that one would be valuable.

\section{Conclusions}
\label{section:conclusion}
When a pandemic-scale cyber incident will occur is unknown, but preparing now will help maximize how we weather its effects. COVID-19 has been catastrophic, even above and beyond the toll on human health. The cybersecurity community should learn every relevant lesson from it and act now to prepare for pandemic-scale cyber events.

The COVID-19 pandemic serves as a stark reminder of the need for robust business continuity plans in the face of unforeseen, high-impact events. Nassim Nicholas Taleb wrote that ``Black Swan logic makes what you don't know far more relevant than what you do know''~\cite{taleb2010black}. ``Consider,'' he said, ``that many Black Swans can be caused and exacerbated [simply] by their being unexpected.'' Similar black swan events in cybersecurity---unpredictable cyberattacks with devastating consequences---could cripple an organization's digital infrastructure and disrupt critical operations. Just as the pandemic exposed vulnerabilities in physical business continuity, lessons learned can be applied to bolster cybersecurity preparedness. By fostering a culture of resilience and prioritizing cyber incident response plans alongside traditional disaster recovery measures, organizations can be better equipped to weather both foreseeable and unforeseen disruptions, be they biological or digital.

In this paper, we highlight critical lessons learned from the COVID-19 pandemic that can be applied to bolster our defenses against future large-scale cyberattacks. Crises occur when something happens that few believe could happen. By recognizing the characteristics of a pandemic-scale cyber event, the cybersecurity community and our partners can proactively develop a response framework built on collaboration, pre-defined roles, and a robust cyber defense playbook. By implementing these actionable steps informed by the COVID-19 experience, we can safeguard critical infrastructure, minimize economic disruptions, and ensure societal resilience in the face of a new kind of global crisis. This framework paves the way for further discussion and collaboration among policymakers, security professionals, and researchers to translate these lessons learned into concrete action plans, ultimately fostering a more secure digital future.

\section*{Acknowledgements}
The authors are grateful to Steve Bellovin, Dan Geer, Yuri Ito, and Gene Spafford as well as the participants in the Ostrom Workshop on Cyber Public Health and the First Workshop on Cyber Public health for their helpful comments and feedback. We thank our NSPW pre-workshop shepherd, Thomas Millar, for useful guidance and recommendations.

\bibliographystyle{acm}
\bibliography{nspw2024}

\end{document}